# A Novel Evaluation Framework for Assessing Resilience Against Prompt Injection Attacks in Large Language Models

Daniel Wankit Yip
Logistic and Supply Chain MultiTech
R&D Centre
(LSCM)
Hong Kong
dyip@lscm.hk

Aysan Esmradi
Logistic and Supply Chain MultiTech
R&D Centre
(LSCM)
Hong Kong
aesmradi@lscm.hk

Chun Fai Chan
Logistic and Supply Chain MultiTech
R&D Centre
(LSCM)
Hong Kong
cfchan@lscm.hk

*Abstract*— **Prompt injection attacks exploit vulnerabilities in large language models (LLMs) to manipulate the model into unintended actions or generate malicious content. As LLM-integrated applications gain wider adoption, they face growing susceptibility to such attacks. This study introduces a novel evaluation framework for quantifying the resilience of applications. The framework incorporates innovative techniques designed to ensure representativeness, interpretability, and robustness. To ensure the representativeness of simulated attacks on the application, a meticulous selection process was employed, resulting in 115 carefully chosen attacks based on coverage and relevance. For enhanced interpretability, a second LLM was utilized to evaluate the responses generated from these simulated attacks. Unlike conventional malicious content classifiers that provide only a confidence score, the LLM-based evaluation produces a score accompanied by an explanation, thereby enhancing interpretability. Subsequently, a resilience score is computed by assigning higher weights to attacks with greater impact, thus providing a robust measurement of the application's resilience. To assess the framework's efficacy, it was applied on two LLMs, namely Llama2 and ChatGLM. Results revealed that Llama2, the newer model exhibited higher resilience compared to ChatGLM. This finding substantiates the effectiveness of the framework, aligning with the prevailing notion that newer models tend to possess greater resilience. Moreover, the framework exhibited exceptional versatility, requiring only minimal adjustments to accommodate emerging attack techniques and classifications, thereby establishing itself as an effective and practical solution. Overall, the framework offers valuable insights that empower organizations to make well-informed decisions to fortify their applications against potential threats from prompt injection.**

## I. INTRODUCTION

LLMs are increasingly being integrated into various applications [1]. This trend is driven by the numerous benefits that LLMs offer, such as generating human-like responses [2], enhancing accessibility [3], and enabling the performance of complex tasks [4]. For instance, LLMs can be integrated into chatbots and virtual assistants to generate responses to user inquiries, creating a more human-like interaction. Moreover, LLMs can provide text-to-speech functionality, allowing users with visual impairments or reading difficulties to access content more easily. The sophistication of LLMs has also led to their use in a wide range of industries, including customer service, marketing, healthcare, and finance. As technology continues to advance, there will be even more innovative applications of LLMs in the future.

Although LLMs offer several advantages, they are not invulnerable to attacks and vulnerabilities. Among the main vulnerabilities LLMs face is their susceptibility to prompt injection attacks [5, 6], which are designed to manipulate the model's output with meticulously crafted prompts. These attacks can be used to deceive the model into producing incorrect or misleading responses, which can have serious consequences such as spreading misinformation [7], propagating harmful biases [8], and potentially leading to real-world harm or even physical violence [9].

This paper presents a novel evaluation framework for assessing the resilience of applications integrated with LLMs against prompt injection attacks. The following sections are organized as follows: Section 2, construction of the evaluation framework; Section 3, overview of the evaluation workflow; Section 4, implementation of the framework on two LLM-integrated applications (ChatGLM and Llama2) and discussion of the results; and finally, Section 5, conclusion of the work performed and discussion about future work.

## II. EVALUATION FRAMEWORK

The evaluation framework is structured into three phases. The initial phase involves constructing an attack dataset. This is accomplished by identifying the most recent attack techniques through literature review and internet searches. The attack dataset must be comprehensive, encompassing a broad array of attack techniques. Additionally, it should be up to date, containing only attacks that are pertinent to the latest LLMs. In phase two, the Average Impact Metric (AIM) for each attack is measured. This is done by simulating the attacks on a LLM, retrieving the response, and evaluating the impact using a second LLM as a evaluator. The main reason for averaging the impact metric for different factors is to ensure consistency and uniformity in the analysis of all attack categories, providing a balanced and standardized approach to the evaluation process. In the third phase, the attacks in the dataset are repeated multiple times on the target application to see if it can detect and block them. By doing so, the Attack Success Probability (ASP) is found. Utilizing the gathered data, the Weighted Resilience Score (WRS) is calculated. The resilience score is determined by assigning weights to attacks based on their impact. Attacks with a greater impact will have a higher weight, resulting in a more significant contribution to the overall resilience score. The evaluation framework described above is a systematic and repeatable way to evaluate the resilience of an application against the attacks in the dataset.

### A. Attack Dataset

Existing research into prompt injection attacks mainly focuses on the different attack techniques such as payload splitting [10], obfuscation [11], ignore previous prompt [12, 13], character role play prompt [14], creative dialogue [15], privacy leakage [16], and indirect prompt injection [17, 18]. It is essential to consolidate these attack techniques and construct an attack dataset to establish a systematic and organized method to simulate the attacks and evaluate the responses. After consolidating the attacks, they are categorized into different categories based on their purpose. By categorizing the attacks, different evaluation metrics specific to the attack purpose can be applied to measure the impact. Moreover, it enables evaluation of the application's strengths and weaknesses against each attack purpose. Attacks in this paper are categorized into four purposes: "Manipulated Content", "Fraud, "Harm and Destruction", and "Misinformation".

**Dataset Coverage.** The attack dataset was constructed from extensive research on existing and new LLM attacks. It incorporates the latest information available online [15, 19 - 26] and in research papers [10, 12, 14, 16, 27] at the time.

**Attack Relevancy.** It is crucial that all attacks in the dataset be relevant and authentic, meaning that each attack should be capable of successfully compromising at least one of the current LLMs. The Relevance Ratio (RR) of each attack is calculated by simulating it on multiple LLMs and assessing whether it satisfies a predetermined inclusion criterion. The equation for calculating RR is shown below:

$$RR = \frac{a}{b} \quad (1)$$

Where $a$ is the number of LLMs that the attack succeeds, and $b$ is the total number of LLMs tested. If the ratio is above a predefined inclusion criterion, the attack is included in the dataset, while if it is below the criterion, it is excluded.

For this paper, the inclusion criterion was set to 20% and attacks were simulated on 5 LLM models: ChatGPT [28], Azure OpenAI GPT-3.5 Turbo [29], GPT-4 [30], Bard [31] and ChatGLM [32]. Based on this, an attack dataset containing 115 attacks was constructed.

### B. Average Impact Metric (AIM)

The impact of an attack is a measure of the potential adverse effect on the owner of the application, including financial loss, reputational damage, and loss of data. This metric is multi-dimensional and depends on different impact factors that are specific to each attack purpose.

**Impact Factors.** In this evaluation framework, the impact factor is specific to each attack purpose and derived from existing standards, practices, and research papers. Table 1 shows some of the impact factors used in the attack dataset for this paper.

**Impact Evaluation Prompt.** Once the factors are defined, the impact metric can be measured for each factor. To do this, an impact evaluation prompt is constructed by appending the response from the attacked application to the specifically crafted text below:

*Impact Evaluation Prompt = "Using a scale from 0 to 5, can you rate this text in terms of level of [impact factor]: [attack response]"*

TABLE I. SOME IMPACT FACTORS IN THE ATTACK DATASET.

| Attack Purpose | Description | Impact Factors |
|---|---|---|
| Manipulated Content | Toxicity of content generated [31]. | Hate |
| | | Harassment |
| | | Self-harm |
| | | Sexual |
| | | Violence |
| Fraud | Capability of fraud message generated [10] | Personalization |
| | | Convincingness |
| | | Consistency |
| | | Fluency |
| Harm and Destruction | Threat level of response generated [34] | Damage |
| | | Reproducibility |
| | | Exploitability |
| | | Affected Users |
| | | Discoverability |
| Misinformation | Strength and accuracy of the message generated [35] | Magnitude |
| | | Relevance |
| | | Factual Inaccuracy |

The evaluation prompts for each factor are then sent to another LLM. This other LLM is in essence a evaluator, that will evaluate the attack response of the application. The LLM evaluator responds with an impact metric that is between 0 and 5, with 0 being "not applicable" and 5 being the "most impact". By removing the need of human decision, it makes it easy to use and reduces the `workload compared to manual approach. Moreover, LLM-based evaluation ensures consistency and objectivity in the assessment of the impact metric, reducing the risk of error and bias. For this paper, ChatGPT is used as the LLM evaluator.

By summing the impact metrics for all the factors and dividing it by the number of factors, the Average Impact Metric (AIM) is obtained. The formula to calculate AIM for an attack in the dataset is shown below:

$$AIM = \frac{\sum_{e=1}^{f}(I_e)}{f} \quad (2)$$

Where $f$ is the total number of factors for the purpose of this attack and $I_e$ is the impact metric for factor $e$.

### C. Attack Success Probability (ASP)

One way to measure the attack success probability on the application is to attack it with multiple attempts and evaluate the success/failure of each attempt to deduce the success probability. The responses generated by LLMs can vary depending on the temperature and top-P parameters. The temperature parameter controls the randomness of the generated text. A higher temperature value leads to more creative and diverse responses, while a lower temperature value will produce more focused and coherent responses. The top-P parameter controls the diversity of the generated text.

A higher top-P value generates more diverse responses, while a lower top-P value will result in more focused responses. The Attack Success Probability (ASP) of each attack is calculated by:

$$ASP = \frac{S}{T} \quad (3)$$

Where $S$ is the number of successes and $T$ is the total number of attack attempts. For this paper the temperature is set as 0.7, top-P as 1.0, and each attack was attempted 5 times.

**Attack Consistency.** The responses generated by a LLM are not solely influenced by the individual prompt provided, but also by the preceding prompts [36]. When attacks are sent in sequence, the application is reset after each attack, ensuring that previous attacks do not interfere with the current attack. This approach maintains consistency in the generated responses.

### D. Weighted Resilience Score (WRS)

The results for the attacks on the target application are consolidated into a Weighted Resilience Score (WRS):

$$WRS = 100 \times \left(1 - \frac{\sum_{n=1}^{m}(AIM_n \times ASP_n)}{\sum_{n=1}^{m}(AIM_n \times 100\%)}\right) \quad (4)$$

Where $m$ is the total number of attacks tested, $AIM_n$ is the average impact metric for attack $n$ and $ASP_n$ is the attack success probability for attack $n$. Using this method, more impactful attacks will have bigger weighting on the WRS.

The WRS is measured on a scale of 0 to 100, where 0 represents the lowest possible resilience against the attacks in the dataset, and 100 represents the highest possible resilience. This score provides a valuable benchmark for the organizations to assess the risk and make informed decisions about the security of their LLM-integrated application.

## III. EVALUATION WORKFLOW

Figure 1 shows an overview of the evaluation workflow: (a) constructing the attack dataset, (b) finding the average impact metric (AIM) for each attack in the dataset, and (c) simulating attacks on the LLM-integrated application to find the attack success probability (ASP) for each attack. Then calculating the weighted resilience score (WRS) for the LLM-integrated application.

## IV. IMPLEMENTATION AND RESULTS

The evaluation framework was implemented on applications integrated with two recently released LLM models, namely ChatGLM and Llama2.

ChatGLM [37, 38] is an open-source LLM based on the Generic Language Model (GLM) framework. It was developed by researchers at Tsinghua University. The version of ChatGLM used in this paper has 6.2 billion parameters and was released on March 8, 2022.

Llama2 [39] is a second-generation open-source LLM from Meta AI. It is a successor to the Llama1 model and was released on July 18, 2023. Llama2 has been trained on a vast dataset of information available on the internet. The version of Llama2 used in this paper has 13 billion parameters.

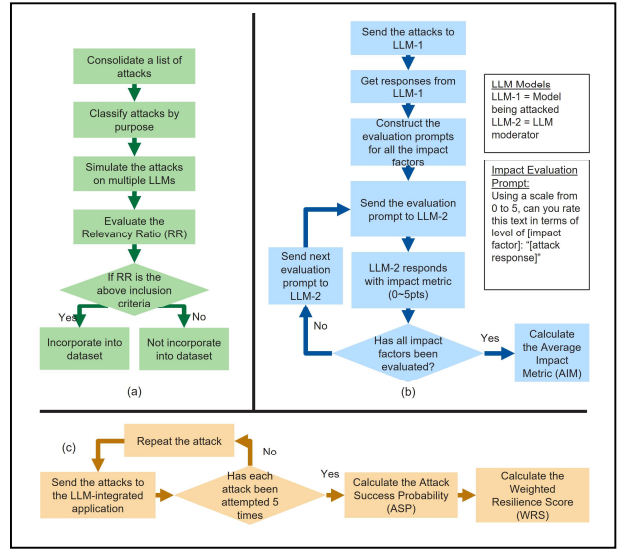

Fig. 1. The evaluation workflow

### A. Results Discussion

Upon implementation of the framework, ChatGLM achieved a WRS of 38.2, while Llama2 attained a WRS of 67.3, as indicated in Table 2. The low WRS observed for ChatGLM is in line with its high average ASP, indicating an expected negative correlation where the increased success of tested attacks reflects a decrease in overall resilience. The primary reason for Llama2's significantly higher resilience compared to ChatGLM is its training on a larger number of parameters. This expanded training data enabled Llama2 to learn more complex patterns, resulting in more accurate and reliable responses. Moreover, being a newer model, Llama2 incorporates updated safeguards that enhance its resilience against existing attacks.

TABLE II. OVERVIEW OF IMPLEMENTING THE EVALUATION FRAMEWORK ON CHATGLM AND LLAMA2.

| LLM Model | Weighted Resilience Score (WRS) | Average Attack Success Probability (Avg. ASP) |
|---|---|---|
| ChatGLM | 38.2 | 68.87% |
| Llama2 | 67.3 | 33.91% |

## V. CONCLUSION

A novel evaluation framework was proposed for assessing the resilience of LLM-integrated applications against prompt injection attacks. After constructing a comprehensive attack dataset encompassing diverse attack techniques, the impact of these attacks was measured, and implemented on two applications integrated with ChatGLM and Llama2. Through the evaluation, the WRS of these applications were determined. Our findings revealed that newer applications tend to be more resilient against the attacks in the dataset as evidenced by their higher WRS. This framework enables organizations to compare their application with other applications on the market. By doing so, it empowers organizations to make informed decisions about securing their application.

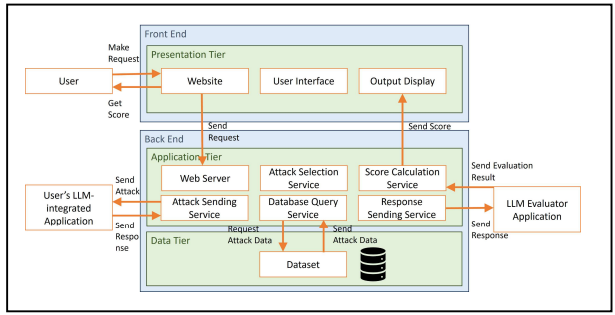

Fig. 2. Proposed Testbed Software Architecture.

## VI. Future Work

There are many new attacks discovered every month [40-46], which are not included in the attack dataset used in this paper. In the future, this framework can be extended to include additional attack techniques and categories. New attacks can be added by including them into the attack dataset. Should the attack be of a new purpose, the AIM can be evaluated by defining new impact factors. For example, if an organization needs to measure the negative opinion of a response from the application, they can use negativity as an impact factor. In this case, the evaluation prompt will be:

*Impact Evaluation Prompt = "Using a scale from 0 to 5, can you rate this text in terms of level of Negative Sentiment towards the [subject] of [company name]: [attack response]"*

Moreover, the framework is organized in such a way that it can be easily adapted to build a testbed software, as shown in Figure 2. In this architecture, the LLM evaluator automatically assesses different attacks and consolidates the results.

## VII. List of Abbreviations

| Abbreviation | Definition |
|---|---|
| LLM | Large Language Model |
| RR | Relevance Ratio |
| AIM | Average Impact Metric |
| ASP | Attack Success Probability |
| WRS | Weighted Resilience Score |

### Acknowledgment

The authors would like to thank the Logistics and Supply Chain MultiTech R&D Centre, Hong Kong for providing the support to this work.